\documentclass[ams,showpacs,preprint]{revtex4}
\usepackage{amsmath}
\usepackage{epsfig}

\newcommand{\re}{\ref}
\newcommand{\be}{\begin{equation}}
\newcommand{\ee}{\end{equation}}

\newcommand{\la}{\label}
\newcommand{\ber}{\begin{eqnarray}}
\newcommand{\eer}{\end{eqnarray}}

\begin{document}

\title{Improved transverse $(e,e')$ response function of  $^3$He at intermediate momentum transfers
}

\author{Victor D. Efros$^{1}$,
  Winfried Leidemann$^{2}$, 
  Giuseppina Orlandini$^{2}$,
  and Edward L. Tomusiak$^{3}$ 
  }

\affiliation{
  $^{1}$Russian Research Centre 
  "Kurchatov Institute",  123182 Moscow,  Russia\\ 
  $^{2}$Dipartimento di Fisica, Universit\`a di Trento, and
  Istituto Nazionale di Fisica Nucleare, Gruppo Collegato di Trento,
  I-38050 Povo, Italy \\
  $^{3}$Department of Physics and Astronomy,
  University of Victoria, Victoria, BC V8P 1A1, Canada\\
}

\begin{abstract}
The transverse electron scattering response function of $^3$He is studied in
the quasi-elastic peak region for momentum transfers between 500 and 700 MeV/c. 
A conventional description of the process leads to results at a substantial variation with
experiment. To improve the results, the present calculation is done in a reference
frame (the  ANB or Active Nucleon Breit frame) which diminishes the
influence of relativistic effects on nuclear states. The laboratory frame response
function is then obtained via a kinematics transformation.
In addition, a one--body nuclear current operator is employed that includes
all leading order relativistic corrections. Multipoles of this operator are listed.  
It is shown that the use of the ANB frame leads to a sizable shift of the quasi-elastic
peak to lower energy and, contrary to the relativistic current, also to an increase of the 
peak height. The additionally considered meson exchange current contribution
is quite small in the peak region. In comparison with experiment one finds an excellent agreement of the peak positions.  
The peak height agrees well with experiment for the lowest considered
momentum transfer (500 MeV/c), but tends to be too high for higher momentum transfer 
(10\% at 700 MeV/c).
\end{abstract}

\bigskip

\pacs{25.30.Fj, 21.45.-v, 21.30.-x}

\maketitle

\section{Introduction}

In Ref. \cite{elot2004}  we studied the longitudinal electron scattering response function
of trinucleons. 
We, as well as others, \cite{elot2004,hann04,bochum} observed that for increasing momentum
momentum transfer $q$, in particular for $q>$ 500 MeV/c, the non-relativistic
theoretical results increasingly deviate from experiment. 
A similar problem arises in the case of the transverse response of 
trinucleons \cite{dmetal,hann04,bochum}.  These problems appear to be related in part
to a deficiency of the non--relativistic nuclear dynamics at such $q$ values.
In Ref. \cite{elot2005} methods were proposed which would allow the extension of such 
non-relativistic calculations to higher $q$. These methods proved to be efficient in
the case of the longitudinal response.

In the present work, with the help of such a method we analyse
the transverse response function of $^3$He  in the quasi-elastic peak region.
Another improvement on the non--relativistic description in the present work
results from our taking into account all the leading
order relativistic corrections to the one-body electromagnetic current operator.
Such corrections 
have been employed in the deuteron case \cite{rgwt97} and they were included
\cite{bochum,hann04} when calculating magnetic form factors for
elastic electron scattering on trinucleons.
However they have not been previously taken into account for the A=3  transverse responses.
Here we account for these corrections via considering the current operator 
that contains all the correction terms of the $M^{-3}$ order.
We calculate this operator proceeding from corresponding matrix elements \cite{rgwt97} of the current.  
 
Our preceding study of the transverse response of $^3$He \cite{dmetal} was done
in the framework of a non--relativistic description with inclusion of full final state 
interaction via  the Lorentz integral transform method
\cite{ELO94,ELOB07}. We used the BonnA NN potential \cite{Bonn} 
plus conventional NNN forces as a nuclear dynamics input. Use of the BonnA potential
gave a "unique" prescription for meson exchange contributions to the electromagnetic current of the nucleus. It is of interest,
however, to use for our present study more modern NN interactions such as the AV18 potential
\cite{AV18}. Our results for the transverse response functions of the trinucleons using
the AV18 NN plus UIX NNN \cite{UrbIX} potentials have recently appeared \cite{FBS09} for the
threshold region.  In that paper we describe our procedure of using the Arenh\"ovel-Schwamb technique
\cite{AS}
meson exchange currents of the AV18 potential.  In the
present work we extend our considerations to the quasi-elastic region and consider various intermediate
momentum transfers.

\section{Formulation}

In the one photon exchange approximation
the cross section for the process of inclusive electron scattering on a nucleus
is given by  
\begin{eqnarray}
{\frac{d^2\sigma} {d\Omega\,d\omega}}\ =\ \sigma_{Mott}\ \bigg[\ {\frac{Q^4}{q_{\rm lab}^4}}\,
R_L(q_{\rm lab},\omega_{\rm lab})\ +
\ \left(\ {\frac{Q^2}{2q_{\rm lab}}^2}+\tan^2{\frac{\theta}{2}}\ \right)\,R_T(q_{\rm lab},\omega_{\rm lab})\bigg]
\la{sigma}\end{eqnarray}
where $R_L$ and $R_T$ are the longitudinal and transverse response functions
respectively, $\omega_{\rm lab}$ is the electron energy loss, $q_{\rm lab}$ is the magnitude of the 
electron momentum transfer, $\theta$ is the electron scattering angle, and 
$Q^2 = q_{\rm lab}^2 - \omega_{\rm lab}^2$.

In the present work we study the transverse response function.  It may be written down as 
\be
R_T(q_{\rm lab},\omega_{\rm lab})={\overline \sum}_{M_i}\sum\!\!\!\!\!\!\!\int\,df
({\bf J}_t^\dag)_{i{\bar f}}\cdot({\bf J}_t)_{{\bar f}i}\delta(E_{\bar f}-E_i-\omega_{\rm lab}). 
\la{resp}\ee
Here the subscripts $i$ and ${\bar f}$ label, respectively, an initial state  and final states, including their  total momenta ${\bf P}_i$ and 
${\bf P}_{\bar f}$. One may write $d{\bar f}=d{\bf P}_{\bar f}df$. Eq. (\re{resp}) contains $df$ only. The notation
$E_{i}$, $E_{{\bar f}}$ refers to total initial and final--state energies. (In \cite{dmetal} the notation $E_{i,f}$ was used
for internal energies.) The quantities $({\bf J}_t)_{{\bar f}\,i}$ are on--shell matrix elements of the transverse component of the nuclear current
operator $\bar{\bf J}({\bf q},\omega)$,
\be ({\bf J}_t)_{{\bar f}i}\delta({\bf P}_{\bar f}-{\bf P}_i-{\bf q})=
\langle\Psi_{{\bar f}}|{\bar{\bf J}}_t({\bf q},\omega)|\Psi_{i}\rangle,
\la{me}\ee 
taken at ${\bf q}={\bf q}_{\rm lab}$, $\omega=\omega_{\rm lab}$, ${\bf P}_i=0$.
The states entering here are eigenstates of the total Hamiltonian with eigenenergies $E_{{\bar f}}$ and $E_{i}$.
They are normalized as
\be
\langle\Psi_{\bar f}|\Psi_{{\bar f}'}\rangle=\delta({\bar f}-{\bar f}'),\qquad
\langle\Psi_i|\Psi_{i'}\rangle=\delta({\bf P}_i-{\bf P}_{i'}).
\ee

The above relationships  refer to the laboratory reference frame.   It is useful also  to consider 
a response--type quantity $R_T^{\rm fr}$ defined by the same relationships
referring to another reference frame. We shall denote the corresponding quantities $q_{\rm fr}$, etc. 
In particular, the states $\Psi_{{\bar f}}$ and $\Psi_{i}$ then will be eigenstates
of the total Hamiltonian in the reference frame considered. For the class of reference frames moving 
with respect to
the laboratory frame along the ${\bf q}$ direction the following relationship is valid:
\be R_T(q_{\rm lab},\omega_{\rm lab})=\frac{E_i^{\rm fr}}{M_T}R_T^{\rm fr}(q_{\rm fr},\omega_{\rm fr}).\la{rere}\ee 
Here $M_T$ is the mass of the target.

Relativistic effects are present in Eq. (\re{me}) both in the states  $\Psi_{i},\Psi_{{\bar f}}$
and in the nuclear current operator. To account for the former effects we proceed
as in the longitudinal case \cite{elot2005} and introduce the active nucleon Breit (ANB)
frame. In the ANB frame,  the nucleus has the momentum $-A{\bf q}_{\rm ANB}/2$ 
in the initial state, ${\bf q}_{\rm ANB}$ being the momentum transfer from the
electron to the nucleus in this reference frame. At high $q$ values,
nucleon momenta in the initial state have the values of about $-{\bf q}_{\rm ANB}/2$ in
this reference frame. 
In the final state in quasi-free kinematics 
the active nucleon has a momentum  about ${\bf q}_{\rm ANB}/2$ while the
momentum of each of the other nucleons remains at about $-{\bf q}_{\rm ANB}/2$.
Thus, typical initial and final state 
nucleon momenta are restricted to magnitudes of about $q_{\rm ANB}/2\simeq q/2$
in the ANB reference frame while, say, in the laboratory frame
nucleon momenta up to $q$ are present. Furthermore,  it also follows  from the above
that  the energy transfer $\omega_{\rm ANB}$ 
in the  ANB reference frame is 
zero at the quasi-elastic peak, and this
applies both to the relativistic and the non--relativistic case. Therefore, 
even when one treats the nucleus non--relativistically
the peak remains at the same position as in the relativistic case.
This contrasts with a description of the process in the laboratory reference frame
where positions of the peak in the relativistic and the non--relativistic cases
would differ considerably. 
Hence non-relativistic calculations in the quasi-elastic region should be done in the ANB
frame to minimize errors due to relativistic effects. The laboratory response function 
sought for is obtained subsequently with the help of Eq. (\re{rere}) with "ANB" being
substituted for "fr".

We perform  the corresponding non--relativistic calculation in the ANB reference frame. One defines  
the internal current operator ${\bf J}$ obtained by taking a matrix element 
in the center of mass subspace of the total current operator:
\be {\bf J}\delta({\bf P}_{\bar f}-{\bf P}_i-{\bf q})=
\langle{\bf P}_{{\bar f}}|{\bar{\bf J}}({\bf q},\omega)|{\bf P}_{i}\rangle.\la{j1}\ee
At ${\bf P}_i=P_i{\hat{\bf q}}$, ${\hat {\bf q}}$ being $q^{-1}{\bf q}$, this operator may be written as ${\bf J}({\bf q},\omega,P_i)$. 
One may then rewrite Eq. (\re{resp}) as 
\be
R_T^{\rm ANB}(q_{\rm ANB},\omega_{\rm ANB})={\overline \sum}_{M_i}\sum\!\!\!\!\!\!\!\int\,df
\langle\psi_i|{\bf J}_t^\dag|\psi_f\rangle\cdot\langle\psi_f|{\bf J}_t|\psi_i\rangle
\delta\left(e_f-e(q_{\rm ANB},\omega_{\rm ANB})\right),
\la{resp1}\ee
where the transverse component ${\bf J}_t$ of ${\bf J}({\bf q},\omega,P_i)$ is used and 
the values ${\bf q}={\bf q}_{\rm ANB}$, $\omega=\omega_{\rm ANB}$, \mbox{$P_i=-Aq_{\rm ANB}/2$} are set.
Here $\psi_i$ and $\psi_f$ are the non--relativistic internal states. They are independent of the center of mass momenta. The energy $e_f$
is the internal energy in the final state, and
\[e(q_{\rm ANB},\omega_{\rm ANB})=e_i+\omega_{\rm ANB}+\frac{\left(P_i^{\rm ANB}\right)^2-\left(P_f^{\rm ANB}\right)^2}{2M_T}
=e_i+\omega_{\rm ANB}+\frac{(A-1)q^2_{\rm ANB}}{2M_T},\]
where $e_i$ is the internal energy in the initial state. One also has 
\[ q_{\rm ANB}=\gamma(q_{\rm lab}-\beta\omega_{\rm lab}),\qquad\omega_{\rm ANB}=\gamma(\omega_{\rm lab}-\beta q_{\rm lab}),\qquad \gamma=(1-\beta^2)^{-1/2},\] 
\[\left[M_T^2+\left(P_i^{\rm ANB}\right)^2\right]^{1/2}=\gamma M_T,
\qquad \beta=\frac{q_{\rm lab}}{2(M_T/A)}\left[1+\frac{\omega_{\rm lab}}{2(M_T/A)}\right]^{-1}.\]
(One gets $\omega_{\rm ANB}=0$ when substituting $\omega_{\rm lab}=\left[(M_T/A)^2+q_{\rm lab}^2\right]^{1/2}-M_T/A$ 
in the expression for $\omega_{\rm ANB}$. This is in agreement with the said above.) 

\section{The nuclear current operator and its multipole decomposition}  

We employ the transition current operator that is a sum of one--body
and two--body currents. In \cite{dmetal} the non--relativistic expression for the one--body current was 
used. For the present applications we have calculated relativistic corrections to the one--body current operator. 
To do this we proceeded from the expressions for matrix elements of the one--body current
of the form $\langle{\bf p}_f|{\bar{\bf J}}|{\bf p}_i\rangle$ listed in Ref. \cite{rgwt97}.
The operator so obtained reproduces  these expressions. 

This operator denoted ${\bf J}^{(1)}$ includes all 
the relativistic corrections up to order $M^{-2}$ 
i.e. in addition to the non--relativistic spin current and convection current terms
 of order $M^{-1}$ it includes all the terms of order $M^{-3}$. 
Our expression for this operator given below is the internal operator as defined
by Eq. (\re{j1}).
We also assume that the initial momentum ${\bf P}_i$ is directed along ${\bf q}$. 
(This is the case for the 
ANB reference frame.) Then the current operator includes dependence on ${\bf q}$, $\omega$ and the magnitude of ${\bf P}_i$.
In the expression for it below  all the momentum operators are placed
on the right hence rendering the operators in non-symmetric forms. Nevertheless the Hermiticity of ${\bf J}({\bf x})$ is still intact but in momentum space reads as
${\bf J}^\dag({\bf q}) ={\bf J}(-{\bf q})$.  
We use the notation ${\bf r}'={\bf r}-{\bf R}$ and ${\bf p}'={\bf p}-A^{-1}{\bf P}$, ${\bf P}$ and ${\bf R}$ being the total
momentum operator and the non--relativistic center of mass operator. 

The resulting one--body current operator is
\ber
{\bf  J}^{(1)}({\bf q},\omega,P_i)={\bf j}_{spin}+{\bf j}_p+{\bf j}_q+\Delta {\bf j}+
(\omega/M){\bf j}_\omega,
\la{cur1}\eer
\ber
{\rm with}\qquad {\bf j}_{spin}=
e^{i{\bf qr}'}\frac{i[{\vec \sigma}\times{\bf q}]}{2M}
\left[G_M\left(1-\frac{q^2}{8M^2}\right)-G_E\frac{\kappa^2q^2}{8M^2}\right],\la{spin}\\ 
{\bf j}_{p}=e^{i{\bf qr}'}\frac{{\bf p}'}{M}\left\{G_E\left[1-
\frac{q^2}{8M^2}(\kappa^2+2)\right]+G_M\frac{q^2}{8M^2}\right\},\la{p}\\
{\bf j}_{q}=e^{i{\bf qr}'}\frac{\kappa{\bf q}}{2M}
\left\{G_E\left[1-\frac{q^2}{8M^2}(\kappa^2+3)\right]+
G_M\frac{q^2}{4M^2}\right\},\la{qq}\\
\Delta {\bf j}=\frac{e^{i{\bf qr}'}}{8M^{3}}\biggl\{-2G_E\left[\kappa{\bf q}(p')^2
+2{\bf p}'(p')^2+2\kappa{\bf p}'({\bf p}'\cdot{\bf q})\right]\biggr.\nonumber\\
+\left[G_M-G_E(1+2\kappa^2)\right]{\bf q}({\bf p}'\cdot{\bf q})\nonumber\\
-2iG_E[{\vec \sigma}\times{\bf q}]\left[(p')^2+\kappa({\bf p}'\cdot{\bf q})\right]
\nonumber\\
\biggl.+i(G_E-G_M)[{\bf p}'\times{\bf q}]\left[\kappa({\vec \sigma}\cdot {\bf q})+
2({\vec \sigma}\cdot {\bf p}')\right]\biggr\},\la{del}\\
{\rm and}\qquad  {\bf j}_\omega=e^{i{\bf qr}'}\frac{G_E-2G_M}{8M}\left({\bf q}+i\kappa
[{\vec \sigma}\times{\bf q}]+2i[{\vec \sigma}\times{\bf p}']\right).
\la{om}\eer
In the above expressions we use the notation
\[G_{E,M}=G_{E,M}^p(Q^2)\frac{1+\tau_z}{2}+G_{E,M}^n(Q^2)\frac{1-\tau_z}{2},\]
where $G_{E,M}^{p,n}$ are the Sachs form factors. We also denote
\be
\kappa=1+2P_i/Aq.\la{kappa}
\ee
(Note that $2{\bf p}+{\bf q}=2{\bf p}'+\kappa{\bf q}$.) The terms ${\bf j}_{p}$ and ${\bf j}_{q}$ together represent the convection current.
The latter longitudinal 
component of this current does not enter the net response. However, this component,
which contains the charge operator, is required
when one  uses an alternative expression (the Siegert form,  see Eq.(20) from Ref. \cite{dmetal})
for electric multipoles based upon the continuity equation. If we chose to
use this  form in the present calculation then
the charge operator with inclusion of the standard Darwin--Foldy and spin--orbit corrections
would be used for calculating electric multipoles. In detail this is given by
\footnote{In \cite{elot2004} the spin--orbit contribution to the charge has been listed
with a misprint. The actual calculation has been performed with the correct
expression.}
\be
{\rho}({\bf q},\omega)=e^{i{\bf qr}'}\left[G_E\left(1-\frac{q^2}{8M^2}\right)
-\frac{G_E-2G_M}{4M^2}i({\vec \sigma}\cdot[{\bf q}\times{\bf p}'])\right].\la{cc}
\ee

For the two--body current operator we use the customary non--relativistic expressions, of the form listed in \cite{dmetal}, Appendix A.
The regularization constants entering the two--body current are adjusted to the NN interaction we use so that the continuity equation is 
satisfied approximately, see \cite{FBS09}.
For high $q$ values the relative contribution of the two--body current in the region of quasi-elastic peak is less
important.

As explained in Ref. \cite{dmetal}, the current operator is to be used in the form of an expansion over the multipole operators $T_{jm}^{\rm el}(q,\omega)$
and $T_{jm}^{\rm mag}(q,\omega)$:
\ber
{\bf J}_t=4\pi\sum_{\lambda={\rm el,mag}}
\sum_{jm}i^{j-\epsilon}T_{jm}^{\lambda}(q,\omega){\bf Y}_{jm}^{\lambda *}({\hat {\bf q}}).
\la{jexp}
\eer
Here $\epsilon=0$ in the electric 
case and $\epsilon=1$ in the magnetic case. 
The quantities  ${\bf Y}_{jm}^{\lambda }$ are electric and magnetic 
vector spherical harmonics 
\cite{varsh}. 
We calculate the multipole operators $T_{jm}^\lambda$ in terms of similar operators $T_{jm}^{l}$ related to the
vector spherical harmonics of the form
\be {\bf Y}_{jm}^l({\hat {\bf q}})=
\sum_{m'+\mu=m}C_{lm'1\mu}^{jm}Y_{lm'}({\hat {\bf q}}){\bf e}_{\mu}. \la{ys}\ee
Here ${\bf e}_\mu$ are the spherical unit basis vectors \cite{varsh}, and 
$l=j\pm 1,j$. From expressing the expansion of Eq. (\re{jexp})  
in terms of the harmonics (\re {ys}) one obtains the  operators
\be T_{jm}^{l}=\frac{1}{4\pi i^{j-\epsilon}}\int d{\bf \hat{q}}
\left({\bf Y}_{jm}^l({\bf \hat{q}})\cdot{\bf J}({\bf q},\omega,P_i)\right).
\la{mtp}
\ee
These operators  are
irreducible tensors of rank $j$.
In accordance with the expressions for the harmonics ${\bf Y}_{jm}^{\rm el,mag}$ in terms of the harmonics 
(\re{ys}) \cite{varsh} one has
\ber
\hat{T}_{jm}^{el}=\left(\frac{j+1}{2j+1}\right)^{1/2}
\hat{T}_{jm}^{j-1}
+\left(\frac{j}{2j+1}\right)^{1/2}\hat{T}_{jm}^{j+1},\la{el}\\
\hat{T}_{jm}^{mag}=\hat{T}_{jm}^{j}.\la{mag}
\eer
Expressions for the components of the multipoles (\re{mtp}) pertaining to the current (\re{cur1}) are listed in the Appendix.
The alternative expression for electric multipoles of the current contains also the multipoles 
\be \rho_{jm}(q)=\frac{1}{4\pi i^j}\int d{\bf \hat{q}} Y_{jm}({\bf \hat{q}})\rho({\bf q},\omega)\la{rhom}\ee
of the charge density operator.
 
The dynamical part of the calculation of the response function  $R_T^{\rm ANB}$ (\re{resp1}) is performed in the same way as for the lab response  
function in \cite{dmetal}.

\section{Results and Discussion}  

As mentioned in the introduction we use the AV18 NN potential and the UIX 3NF
as nuclear force. The calculation is carried out in the ANB frame for eight
momentum transfers $q_{\rm ANB}$: 400, 450, 500, 550, 600, 650, 700, and 750 MeV/c.
We consider electric and magnetic multipole contributions up to a maximal total angular
momentum $J_f^{\rm max}$ of the final state such that a convergent result of $R_T$ is 
obtained for any $q$ value. For instance we take $J_f^{\rm max}=19/2$ and 37/2 for $q=400$ and $q=750$ MeV/c,
respectively. As already poined out before we use the LIT formalism \cite{ELO94,ELOB07} in order to take into account
final state interaction. For the LIT parameter $\sigma_I$ we choose two different values,
namely $\sigma_{I,1}=5$ MeV  and  $\sigma_{I,2}=50$ MeV. We combine both results in the following way
\begin{equation}
L_{tot}(\sigma_R,\sigma_I) =  L(\sigma_R,\sigma_{I,1}) f(\sigma_R) +
        \bigg({\frac{\sigma_{I,2}}{\sigma_{I,1}}}\bigg)^2 L(\sigma_R,\sigma_{I,2}) (1-f(\sigma_R)) \,,
\end{equation}
where $L$ denotes the Lorentz transforms of the response, and
\begin{equation}
f(\sigma_R) = \exp(-(\sigma_R/\sigma_0)^6) \,\, (\sigma_R \ge 0) \,\,\,\, {\rm and} \,\,\,\, f(\sigma_R)=1  \,\, (\sigma_R \le 0)
\end{equation}
with $\sigma_0=100$ MeV. This choice has the advantage that one has a relatively large resolution
for the $R_T$ behavior at lower energies, while for the high-energy behavior a smaller resolution
is completely sufficient. The integral equation that corresponds to the transform $L_{tot}$ was solved to extract $R_T$. 
The inversion of the LIT \cite{ELO99,Andreasi05,BELO09} has been made as
described in \cite{dmetal}. 

In Fig.~1 we show $R_T(q_{\rm ANB},\omega_{\rm ANB})$ for the above mentioned eight $q$ values
using our full current operator (relativistic one-body + isovector MEC consistent with AV18). One sees
that the $q_{\rm ANB}$ dependence of $R_T$ exhibits a very regular and smooth pattern.
This allows us to use a spline interpolation to determine $R_T(q_{\rm ANB},\omega_{\rm ANB})$ for intermediate $q_{\rm ANB}$ 
values. In this way we are able to obtain results for $R_T(q_{\rm lab},\omega_{\rm lab})$ via the transformation of
Eq.(\re{rere}) for 500 MeV/c $\le q_{\rm lab} \le 700$ MeV/c.

In the following we investigate three different theoretical aspects:
(i) comparison of lab and ANB frame calculations, 
(ii) relativistic contributions to the one-body current operator, and
(iii) the MEC contribution. We first turn to the comparison of lab and ANB frame results.
In Fig.~2 we show $R_T(q_{\rm lab},\omega_{\rm lab})$ evaluated with the nonrelativistic
one-body current for lab and ANB frame calculations. The ANB results show a sizable shift
of the peak position to lower energies, which grows with increasing $q$. In detail one
has the following shifts, 8.7, 16.7, and 29.3 MeV at $q=500$, 600, and 700 MeV/c. The size
of the shifts is very similar to those found for the longitudinal response function $R_L$
in \cite{elot2005} and corresponds to the 
differences of non-relativistic and relativistic kinetic energies of a nucleon with momentum $q_{lab}$ (see discussion of peak position in section III). 
One also finds an increase of the peak heights, namely by 5.6\%, 10.3\%,
and 16.7\%. The relativistic contribution to the one-body current is illustrated in Fig.~3.
It leads to a reduction of the peak heights of 6.2\%, 8.5\%, and 11.3\% at $q=500$, 600,
and 700 MeV/c, while there are no sizable effects on the peak position. Finally Fig.~4 shows the
MEC contributions. As one might expect they are rather small and decrease with increasing $q$.
In detail one has increases of 3.2\%, 2.7\%, and 2.2\%  for the three considered $q$ values.

Now we turn to a comparison with experimental data (see Fig.~5).
For all the three considered momentum transfers one finds an excellent agreement
of experimental and theoretical peak positions. For $q=500$ MeV/c one also has an
excellent agreement of the peak height. At $q=600$ and 700 MeV/c the theoretical
peak height overestimates the data by about 5\% and 10\%, respectively.
We would like to mention that a different choice for the nucleon form factor fits
should lead to rather small effects only. The reason is that at higher momentum 
transfer $R_T$ is dominated by the spin current contribution where the magnetic nucleon form 
factors enter which for the various fits are rather similar in the  range 500 MeV/c 
$\le$ $q$ $\le$ 700 MeV/c (e.g., compare the dipole fits with those from \cite{MMD}). 
In the present work we do not consider any $\Delta$ degrees of freedom. As shown in \cite{hann04}, 
up to $q=500$ MeV/c there are only tiny $\Delta$ effects in the quasi-elastic region. Also at higher 
$q$ one may expect that the quasi-elastic response is not affected much by $\Delta$ isobar currents 
(compare to deuteron electrodisintegration results, see e.g. \cite{ALT05}). 
The increasing difference between theory and experiment with growing
momentum transfer suggests that unincluded relativistic effects
(wave function boost, dynamical effects) are increasing in importance.
In future we will investigate to see if we can get a better understanding of these
effects in order to improve the comparison with experiment.

For the comparison with the experimental data of Fig.~5 one has to consider that pion production is not 
taken into account in our calculation. The pion production thresholds are at about 180, 200 and 220 MeV at 
$q=500$, 600, and 700 MeV/c, respectively. For $q=500$ MeV/c one can nicely see that the theoretical $R_T$ 
starts to underestimate the experimental $R_T$ in the pion threshold region.

To sum up we can say the following. We have calculated the $^3$He transverse response function $R_T(q,\omega)$
with a realistic nuclear force (AV18 two-nucleon and UIX three-nucleon potential) in the quasi-elastic region 
at 500 MeV/c $\le q \le$ 700 MeV/c with full inclusion of final state interaction. The calculation is carried 
out in the ANB frame with a subsequent transformation of $R_T$ to the lab system. Relativistic effects to the 
one-body current operator as well as meson exchange currents are taken into account. The relativistic effects 
reduce the quasi-elastic peak, while the MEC contributions are rather unimportant. The use of the ANB frame provides
excellent agreement with experimental peak positions. Concerning the peak heigts one finds a good agreement of 
theoretical and experimental results at $q=500$ MeV/c, while theory overestimates data up to 10\% at higher $q$.

\section{Acknowledgement}
Acknowledgements of financial support are given to
the RFBR, grant 07-02-01222-a and RMES, grant NS-3004.2008.2 (V.D.E.),
and to the National Science and Engineering Research Council of Canada (E.L.T.).

\appendix

\section{Multipoles of the one--body current  and  charge operators}
  
In the formulae below we use the notation
\[ \psi_j=j_j(qr')Y_{jm}({\hat {\bf r}}'),\qquad \Pi_a=\sqrt{2a+1},\qquad \Pi_{ab}=
\sqrt{(2a+1)(2b+1)}.\]
The quantity $\partial'_\mu$ below is defined by the relationship $-i\partial'_\mu=p'_\mu$, and
$X_{\gamma\mu}=(\partial'\otimes\partial')_{\gamma\mu}$.
Denoting $-i{\vec\partial}'^{(A)}={\bf p}_A'$ where ${\bf p}_A'$ is the last particle internal
momentum, one has
\[\partial'^{(A)}_\mu=
\left[\frac{A-1}{A}\right]^{1/2}\frac{\partial}{\partial\xi_{A-1,\mu}}.\]
Here the derivative is taken with respect to a component of the last Jacobi vector
defined as 
$\vec{\xi}_{A-1}=\sqrt{(A-1)/A}\,\,
[{\bf r}_A-(A-1)^{-1}\sum_{i=1}^{A-1}{\bf r}_i]$.

Various operators entering the current (\re{cur1}) 
give the following contributions
to the multipoles (\re{mtp}):
\ber
\left(4\pi i^{j-1}\right)^{-1}\int d{\hat{\bf q}}e^{i{\bf qr}'}
i\left({\bf Y}_{jm}^j({\hat{\bf q}})\cdot 
[{\vec \sigma}\times{\hat{\bf q}}]\right)\nonumber\\
=\left(\frac{j}{2j+1}\right)^{1/2}
(\psi_{j+1}\otimes \sigma)_{jm}
-\left(\frac{j+1}{2j+1}\right)^{1/2}
(\psi_{j-1}\otimes\sigma)_{jm},
\eer
\be
\left(4\pi i^{j}\right)^{-1}\int d{\hat{\bf q}}e^{i{\bf qr}'}
i\left({\bf Y}_{jm}^{j+1}({\hat{\bf q}})\cdot 
[{\vec \sigma}\times{\hat{\bf q}}]\right)
=-\left(\frac{j}{2j+1}\right)^{1/2}
\left(\psi_{j}\otimes\sigma\right)_{jm},
\ee
\be
\left(4\pi i^{j}\right)^{-1}\int d{\hat{\bf q}}e^{i{\bf qr}'}
i\left({\bf Y}_{jm}^{j-1}({\hat{\bf q}})\cdot 
[{\vec \sigma}\times{\hat{\bf q}}]\right)
=-\left(\frac{j+1}{2j+1}\right)^{1/2}
\left(\psi_{j}\otimes\sigma\right)_{jm}.
\ee

\medskip

\be
\left(4\pi i^{j-1}\right)^{-1}\int d{\hat{\bf q}}e^{i{\bf qr}'}
\left({\bf Y}_{jm}^j({\hat{\bf q}})\cdot {\bf p}'\right)
=\left(\psi_{j}\otimes\partial'\right)_{jm},
\ee
\be
\left(4\pi i^{j}\right)^{-1}\int d{\hat{\bf q}}e^{i{\bf qr}'}
\left({\bf Y}_{jm}^{j\pm1}({\hat{\bf q}})\cdot {\bf p}'\right)
=\pm \left(\psi_{j\pm1}\otimes\partial'\right)_{jm}.
\ee

\medskip

\be
\left(4\pi i^{j-1}\right)^{-1}\int d{\hat{\bf q}}e^{i{\bf qr}'}
\left({\bf Y}_{jm}^j({\hat{\bf q}})\cdot {\hat{\bf q}}\right)=0,
\ee
\be
\left(4\pi i^{j}\right)^{-1}\int d{\hat{\bf q}}e^{i{\bf qr}'}
\left({\bf Y}_{jm}^{j+1}({\hat{\bf q}})\cdot {\hat{\bf q}}\right)
=-\left(\frac{j+1}{2j+1}\right)^{1/2}\psi_{jm},
\ee
\be
\left(4\pi i^{j}\right)^{-1}\int d{\hat{\bf q}}e^{i{\bf qr}'}
\left({\bf Y}_{jm}^{j-1}({\hat{\bf q}})\cdot {\hat{\bf q}}\right)
=\left(\frac{j}{2j+1}\right)^{1/2}\psi_{jm}.
\ee

\medskip

\ber
\left(4\pi i^{j-1}\right)^{-1}\int d{\hat{\bf q}}e^{i{\bf qr}'}
\left({\bf Y}_{jm}^j({\hat{\bf q}})\cdot{\bf p}'\right) 
({\bf p}'\cdot {\hat{\bf q}})\nonumber\\
=\sqrt{5j}\left\{\begin{array}{ccc}
1&1&2\\j-1&j&j\end{array}\right\}
\left(\psi_{j-1}\otimes X_2\right)_{jm}
+\sqrt{5(j+1)}
\left\{\begin{array}{ccc}
1&1&2\\j+1&j&j\end{array}\right\}
\left(\psi_{j+1}\otimes X_2\right)_{jm},
\eer
\be
\left(4\pi i^{j}\right)^{-1}\int d{\hat{\bf q}}e^{i{\bf qr}'}
\left({\bf Y}_{jm}^{j+1}({\hat{\bf q}})\cdot{\bf p}'\right) 
({\bf p}'\cdot {\hat{\bf q}})=-S^{j+2}_j,
\ee
\be
\left(4\pi i^{j}\right)^{-1}\int d{\hat{\bf q}}e^{i{\bf qr}'}
\left({\bf Y}_{jm}^{j-1}({\hat{\bf q}})\cdot{\bf p}'\right) 
({\bf p}'\cdot {\hat{\bf q}})=S^j_j,
\ee
\ber
S^\lambda_j=\sqrt{\lambda-1}\sum_{\gamma=0,2}
\Pi_\gamma\left\{\begin{array}{ccc}
1&1&\gamma\\\lambda-2&j&\lambda-1\end{array}\right\}
\left(\psi_{\lambda-2}\otimes X_\gamma\right)_{jm}\nonumber\\
+\sqrt{\lambda}\sum_{\gamma=0,2}\Pi_\gamma
\left\{\begin{array}{ccc}
1&1&\gamma\\\lambda&j&\lambda-1\end{array}\right\}
\left(\psi_{\lambda}\otimes X_\gamma\right)_{jm}.
\eer

\medskip

\be
\left(4\pi i^{j-1}\right)^{-1}\int d{\hat{\bf q}}e^{i{\bf qr}'}
\left({\bf Y}_{jm}^{j}({\hat{\bf q}})\cdot{\hat{\bf q}}\right)
\left({\bf p}'\cdot {\hat{\bf q}}\right)=0,
\ee
\be
\left(4\pi i^{j}\right)^{-1}\int d{\hat{\bf q}}e^{i{\bf qr}'}
\left({\bf Y}_{jm}^{j+1}({\hat{\bf q}})\cdot{\hat{\bf q}}\right)
\left({\bf p}'\cdot {\hat{\bf q}}\right)=\sqrt{j+1}\,S,
\ee
\be
\left(4\pi i^{j}\right)^{-1}\int d{\hat{\bf q}}e^{i{\bf qr}'}
\left({\bf Y}_{jm}^{j-1}({\hat{\bf q}})\cdot{\hat{\bf q}}\right)
\left({\bf p}'\cdot {\hat{\bf q}}\right)=-\sqrt{j}\,S,
\ee
\be
S=(2j+1)^{-1}\left[\sqrt{j}(\psi_{j-1}\otimes\partial')_{jm}+
\sqrt{j+1}(\psi_{j+1}\otimes\partial')_{jm}
\right].
\ee

\medskip

\ber
\left(4\pi i^{j-1}\right)^{-1}\int d{\hat{\bf q}}e^{i{\bf qr}'}i
\left({\bf Y}_{jm}^{j}({\hat{\bf q}})\cdot
[{\vec \sigma}\times{\hat{\bf q}}]\right)({\bf p}'\cdot{\hat{\bf q}})
=\left[\frac{(j+1)}{(2j-1)(2j+1)}\right]^{1/2}\nonumber\\\times\left[\sqrt{j-1}
\left(\left(\psi_{j-2}\otimes\partial'\right)_{j-1}\otimes\sigma
\right)_{jm}
+\sqrt{j}\left(\left(\psi_{j}\otimes\partial'\right)_{j-1}\otimes\sigma
\right)_{jm}\right]\nonumber\\
-\left[\frac{j}{(2j+1)(2j+3)}\right]^{1/2}\left[\sqrt{j+1}
\left(\left(\psi_{j}\otimes\partial'\right)_{j+1}\otimes\sigma
\right)_{jm}\right.\nonumber\\
\left.+\sqrt{j+2}\left(\left(\psi_{j+2}\otimes\partial'
\right)_{j+1}\otimes\sigma\right)_{jm}\right].
\eer
\be
\left(4\pi i^{j}\right)^{-1}\int d{\hat{\bf q}}e^{i{\bf qr}'}i
\left({\bf Y}_{jm}^{j+1}({\hat{\bf q}})\cdot
[{\vec \sigma}\times{\hat{\bf q}}]\right)({\bf p}'\cdot{\hat{\bf q}})=\sqrt{j}\,S,
\ee
\be
\left(4\pi i^{j}\right)^{-1}\int d{\hat{\bf q}}e^{i{\bf qr}'}i
\left({\bf Y}_{jm}^{j-1}({\hat{\bf q}})\cdot
[{\vec \sigma}\times{\hat{\bf q}}]\right)({\bf p}'\cdot{\hat{\bf q}})=
\sqrt{j+1}\,S,
\ee
\ber
S=(2j+1)^{-1}\left[\sqrt{j}\left(\left
(\psi_{j-1}\otimes\partial'\right)_{j}\otimes\sigma\right)_{jm}
+\sqrt{j+1}\left(\left(\psi_{j+1}\otimes\partial'
\right)_{j}\otimes\sigma\right)_{jm}\right].
\eer

\medskip

\ber
\left(4\pi i^{j-1}\right)\int d{\hat{\bf q}}e^{i{\bf qr}'}
i\left({\bf Y}_{jm}^{j}({\hat{\bf q}})\cdot[{\bf p}'\times{\hat{\bf q}}]
\right)({\vec \sigma}\cdot{\hat{\bf q}})
=\left[\frac{(j-1)(j+1)}{2j+1}\right]^{1/2}
\nonumber\\
\times\Pi_{j-1}\left\{\begin{array}{ccc}
1&j-2&j-1\\1&j&j-1\end{array}\right\}\left
(\left(\psi_{j-2}\otimes\partial'\right)_{j-1}\otimes\sigma\right)_{jm}
-\left[\frac{j(j+1)}{2j+1}\right]^{1/2}\nonumber\\
\times\sum_{l=j\pm1,j}(-1)^{l-j}\Pi_l
\left[\left\{\begin{array}{ccc}
1&j&j-1\\1&j&l\end{array}\right\}-\left\{\begin{array}{ccc}
1&j&j+1\\1&j&l\end{array}\right\}\right]
\left(\left(\psi_{j}\otimes\partial'\right)_l\otimes\sigma\right)_{jm}
\nonumber\\
-\left[\frac{j(j+2)}{2j+1}\right]^{1/2}
\Pi_{j+1}
\left\{\begin{array}{ccc}
1&j+2&j+1\\1&j&j+1\end{array}\right\}\left
(\left(\psi_{j+2}\otimes\partial'\right)_{j+1}\otimes\sigma\right)_{jm},
\eer
\be
\left(4\pi i^{j}\right)^{-1}\int d{\hat{\bf q}}e^{i{\bf qr}'}i
\left({\bf Y}_{jm}^{j+1}({\hat{\bf q}})\cdot[{\bf p}'\times{\hat{\bf q}}]
\right)({\vec \sigma}\cdot{\hat{\bf q}})=\sqrt{j}\,S,
\ee
\be
\left(4\pi i^{j}\right)^{-1}\int d{\hat{\bf q}}e^{i{\bf qr}'}i
\left({\bf Y}_{jm}^{j-1}({\hat{\bf q}})\cdot[{\bf p}'\times{\hat{\bf q}}]
\right)({\vec \sigma}\cdot{\hat{\bf q}})=\sqrt{j+1}\,S,
\ee
\ber
S=\left(\frac{j}{2j+1}\right)^{1/2}
\left.\sum_{l=j-1,j}(-1)^{l-j}\Pi_l\left\{\begin{array}{ccc}
1&j-1&j\\1&j&l\end{array}\right\}\left
(\left(\psi_{j-1}\otimes\partial'\right)_l\otimes\sigma\right)_{jm}
\right.\nonumber\\
+\left(\frac{j+1}{2j+1}\right)^{1/2}\sum_{l=j,j+1}(-1)^{l-j}\Pi_l
\left\{\begin{array}{ccc}
1&j+1&j\\1&j&l\end{array}\right\}\left
(\left
(\psi_{j+1}\otimes\partial'\right)_l\otimes\sigma\right)_{jm}.
\eer

\medskip

\ber
\left(4\pi i^{j-1}\right)^{-1}\int d{\hat{\bf q}}e^{i{\bf qr}'}i
\left({\bf Y}_{jm}^j({\hat{\bf q}})\cdot
[{\bf p}'\times {\hat{\bf q}}]\right) ({\vec \sigma}\cdot{\bf p}')\nonumber\\
=-\left(\frac{j+1}{2j+1}\right)^{1/2}
\sum_{\gamma=0,2}\sum_{l}\Pi_{l\gamma}\left\{\begin{array}{ccc}
1&1&\gamma\\l&j-1&j\end{array}\right\}
\left(\left(\psi_{j-1}\otimes X_\gamma\right)_l
\otimes\sigma\right)_{jm}\nonumber\\
+\left(\frac{j}{2j+1}\right)^{1/2}
\sum_{\gamma=0,2}\sum_{l}\Pi_{l\gamma}\left\{\begin{array}{ccc}
1&1&\gamma\\l&j+1&j\end{array}\right\}\left(\left(\psi_{j+1}
\otimes X_\gamma\right)_l\otimes\sigma\right)_{jm},
\eer
\ber
\left(4\pi i^{j}\right)^{-1}\int d{\hat{\bf q}}e^{i{\bf qr}'}
i\left({\bf Y}_{jm}^{j+1}({\hat{\bf q}})\cdot
[{\bf p}'\times {\hat{\bf q}}]\right) 
({\vec \sigma}\cdot{\bf p}')=\left(\frac{j}{2j+1}\right)^{1/2}S,\nonumber\\
\left(4\pi i^{j}\right)^{-1}\int d{\hat{\bf q}}e^{i{\bf qr}'}
i\left({\bf Y}_{jm}^{j-1}({\hat{\bf q}})\cdot
[{\bf p}'\times {\hat{\bf q}}]\right) 
({\vec \sigma}\cdot{\bf p}')=\left(\frac{j+1}{2j+1}\right)^{1/2}S,\nonumber\\
S=\sum_{\gamma=0,2}\sum_{l}\Pi_{l\gamma}
\left\{\begin{array}{ccc}
1&1&\gamma\\l&j&j\end{array}\right\}
\left(\left
(\psi_{j}\otimes X_\gamma\right)_l\otimes\sigma\right)_{jm}.
\eer

\medskip

\ber
\left(4\pi i^{j-1}\right)^{-1}\int d{\hat{\bf q}}e^{i{\bf qr}'}i
\left({\bf Y}_{jm}^j({\hat{\bf q}})\cdot[{\vec \sigma}\times{\bf p}']\right)
\nonumber\\
=-\sqrt{6}\sum_{l=j\pm1,j}\Pi_l\left\{\begin{array}{ccc}
1&1&1\\j&j&l\end{array}\right\}
\left(\left(\psi_j\otimes\partial'\right)_l\otimes\sigma\right)_{jm}.
\eer
\ber
\left(4\pi i^{j}\right)^{-1}\int d{\hat{\bf q}}e^{i{\bf qr}'}i
\left({\bf Y}_{jm}^{j\pm1}({\hat{\bf q}})\cdot[{\vec \sigma}\times{\bf p}']\right)
\nonumber\\
=\pm\sqrt{6}\sum_{l}\Pi_l\left\{\begin{array}{ccc}
1&1&1\\j\pm1&j&l\end{array}\right\}
\left(\left(\psi_{j\pm1}\otimes\partial'\right)_l\otimes\sigma\right)_{jm}.
\eer

At deriving these formulae the expressions for ${\bf n}\cdot{\bf Y}_{jm}^l({\bf n})$
in terms of spherical harmonics, and for ${\bf n}Y_{jm}({\bf n})$ and $[{\bf n}\times{\bf Y}_{jm}^l({\bf n})]$
in terms of vector spherical harmonics \cite{varsh} have been used.  We used also the relationship
\[ ({\bf n}\cdot{\bf a})({\bf Y}_{jm}^l({\bf n})\cdot{\bf b})=\left(\frac{l}{2l+1}\right)^{1/2}
\left(\left(Y_{l-1}({\bf n})\otimes a\right)_l\otimes b\right)_{jm}
-\left(\frac{l+1}{2l+1}\right)^{1/2}
\left(\left(Y_{l+1}({\bf n})\otimes a\right)_l\otimes b\right)_{jm}.\]

The spin--orbit component  of the charge density operator (\re{cc}) in (\re{rhom})
leads to the multipoles
\ber \left(4\pi i^j\right)^{-1}\int d{\hat{\bf q}}e^{i{\bf qr}'}
Y_{jm}({\hat{\bf q}})i({\vec\sigma}\cdot[{\hat{\bf q}}\times {\bf p}'])\nonumber\\
=\sqrt{6}\sum_l\Pi_l\left[\left(\frac{j}{2j+1}\right)^{1/2}\left\{\begin{array}{ccc}
1&1&1\\j+1&j&l\end{array}\right\}\left((\psi_{j+1}\otimes\partial')_l\otimes\sigma\right)_{jm}\right].
\eer

\newpage

\begin{figure}[ht]
\centerline{\resizebox*{18cm}{12cm}{\includegraphics*[angle=0]{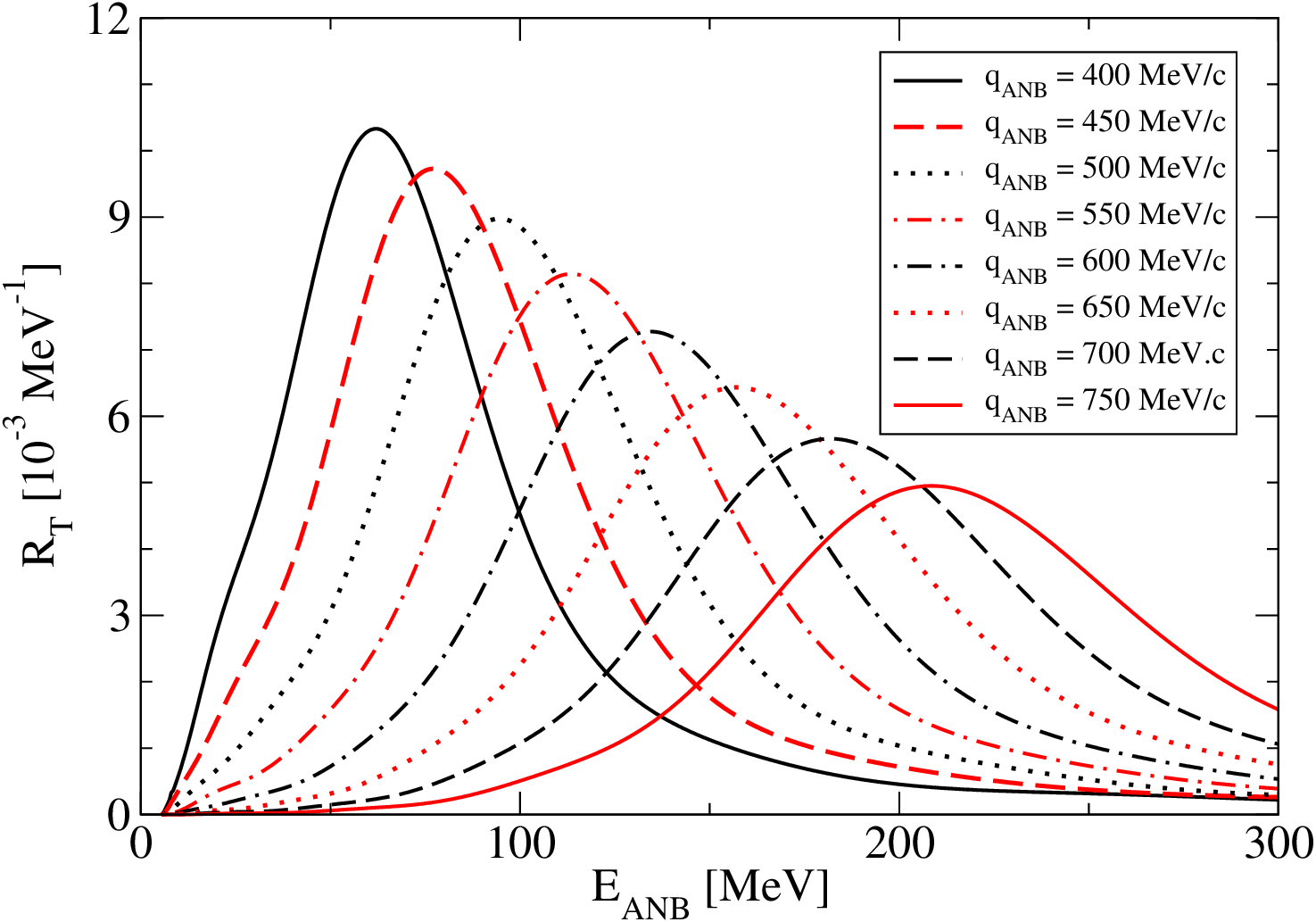}}}
\caption{(color online) $R_T(q_{\rm ANB},E_{\rm ANB})$ of $^3$He with relativistic one-body and meson exchange current at
various $q$ values (internal excitation energy $E_{\rm ANB}=\omega_{\rm ANB}+q^2_{\rm ANB}
/M_T)$.}
\end{figure}

\begin{figure}[ht]
\centerline{\resizebox*{15.cm}{18.cm}{\includegraphics*[angle=0]{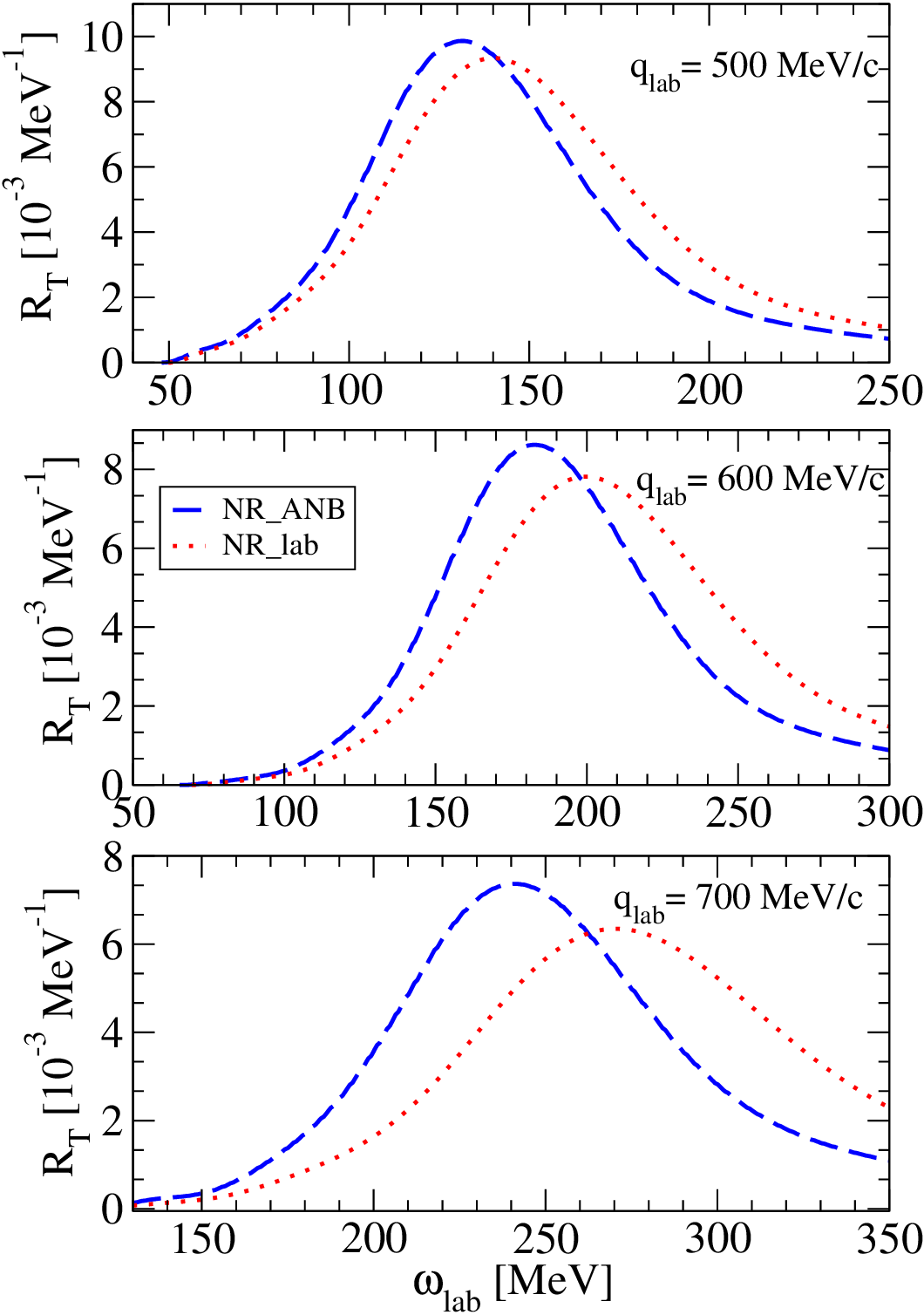}}}
\caption{(color online) $R_T(q_{\rm lab},\omega_{\rm lab})$ of $^3$He from ANB (dashed) and lab (dotted) frame calculations with non-relativistic one-body current.}
\end{figure}

\begin{figure}[ht]
\centerline{\resizebox*{15.cm}{18.cm}{\includegraphics*[angle=0]{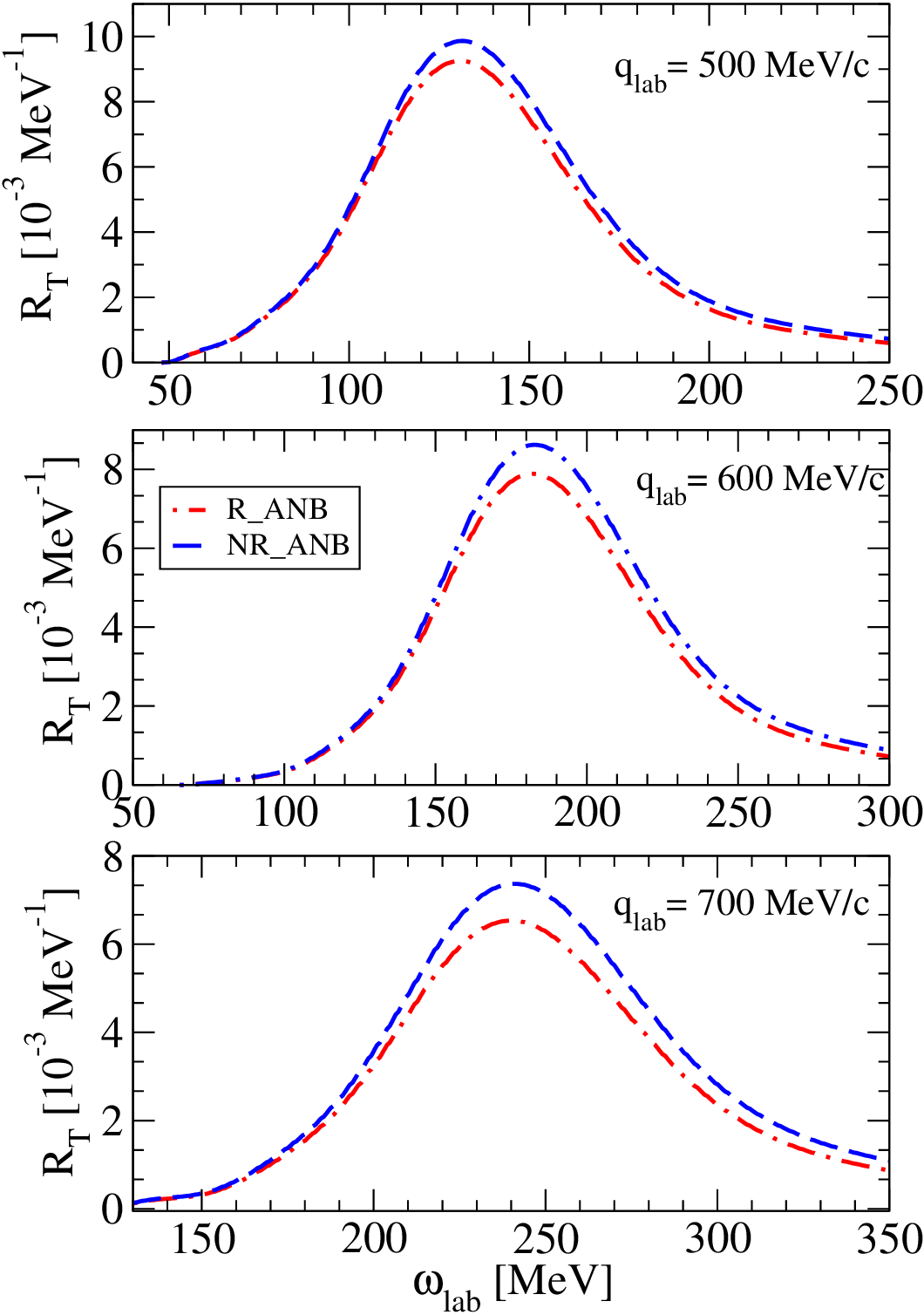}}}
\caption{(color online) $R_T(q_{\rm lab},\omega_{\rm lab})$ of $^3$He from ANB frame calculation with relativistic (dash-dotted)
and non-relativistic (dashed) one-body current.}
\end{figure}

\begin{figure}[ht]
\centerline{\resizebox*{15.cm}{18.cm}{\includegraphics*[angle=0]{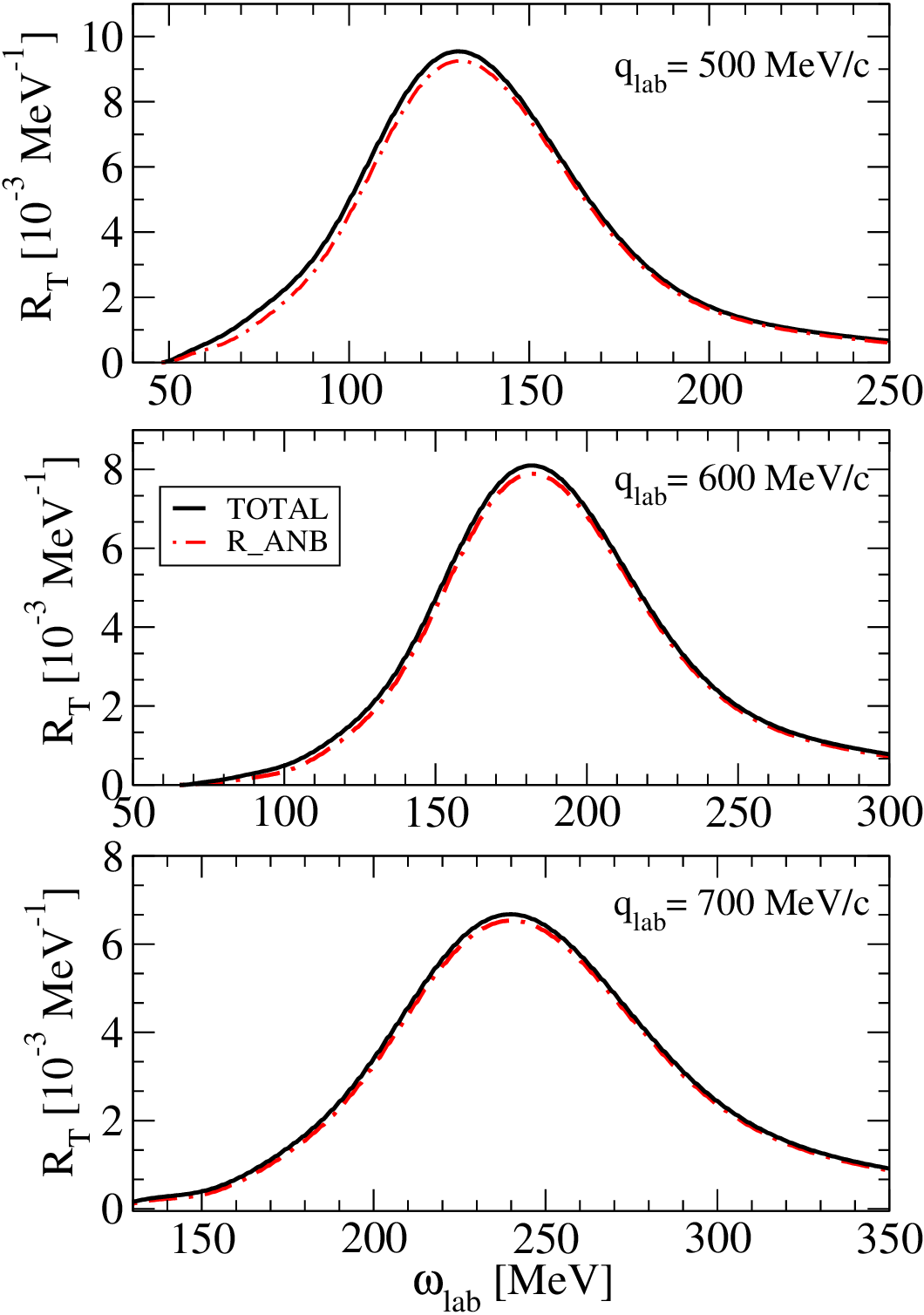}}}
\caption{(color online) $R_T(q_{\rm lab},\omega_{\rm lab})$ of $^3$He from ANB frame calculation with relativistic one-body current with (full) and without (dash-dotted) meson exchange current.}
\end{figure}

\begin{figure}[ht]
\centerline{\resizebox*{15.cm}{18.cm}{\includegraphics*[angle=0]{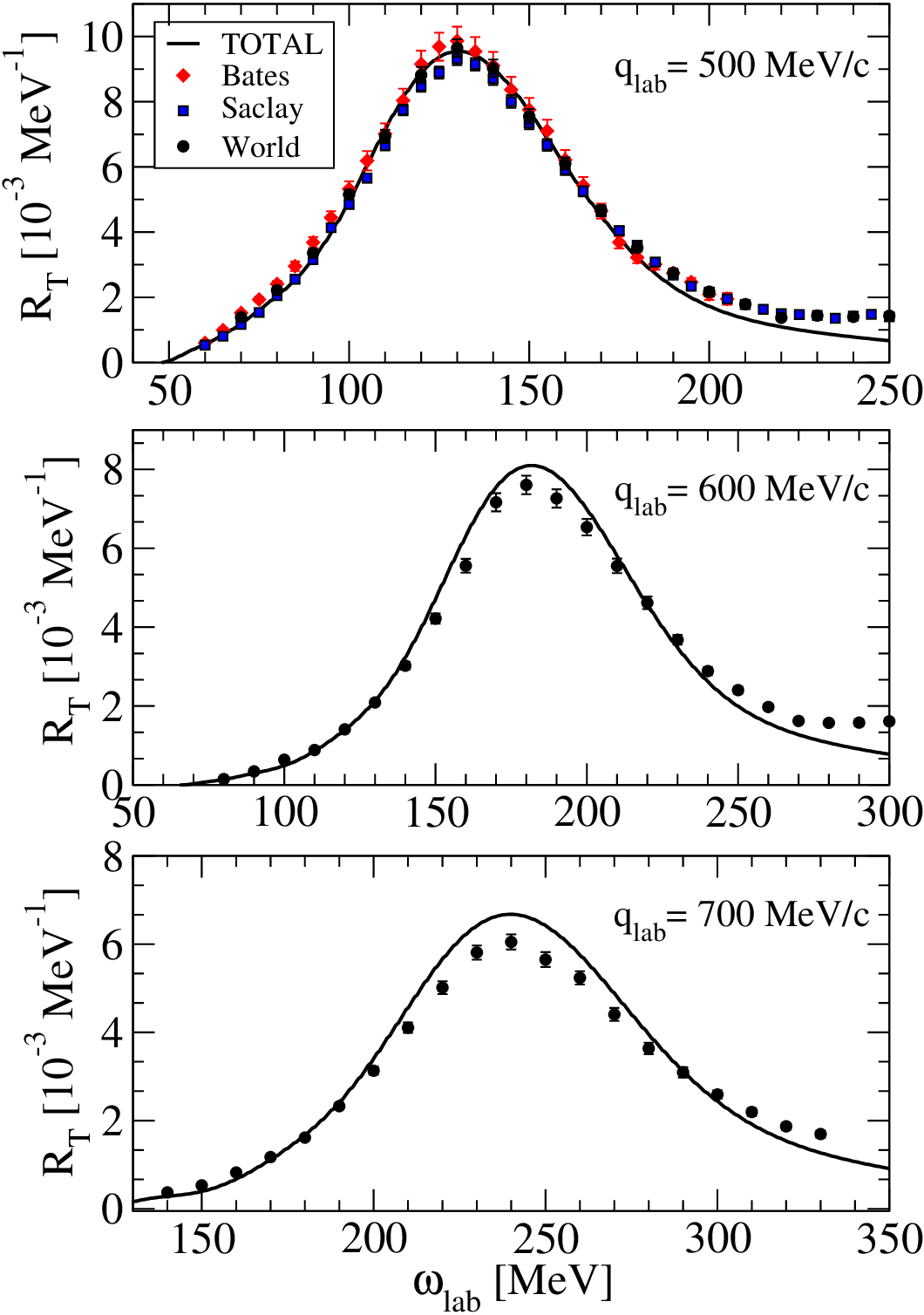}}}
\caption{(color online) $R_T(q_{\rm lab},\omega_{\rm lab})$ of $^3$He from ANB frame calculation with relativistic one-body and meson exchange current (full) in comparison to experimental results from \cite{Saclay} (squares), \cite{Bates} (diamonds),
\cite{world} (circles).}
\end{figure}

\end{document}